\documentclass[twocolumn,floatfix,preprintnumbers,nofootinbib,superscriptaddress]{revtex4-2}

\usepackage{ulem}
\usepackage{bm}
\usepackage{times}
\usepackage{amssymb,amsbsy,amsmath,amsfonts}
\usepackage{graphicx}
\usepackage{float}
\usepackage{color}
\usepackage{morefloats}
\usepackage{rotating}
\usepackage{srcltx}
\usepackage{slashed}
\usepackage{subfigure}
\usepackage{multirow}
\usepackage{verbatim}
\usepackage{hyperref}
\usepackage{tabularx}
\usepackage{braket}
\usepackage{diagbox}
\usepackage{appendix}
\usepackage{makecell}
\usepackage{booktabs}
\usepackage[section]{placeins}

\usepackage[outdir=./]{epstopdf}

\begin{document}
\title{Radiative decays of the $\Omega(2012)$ as a hadronic molecule}
\date{\today}
\author{Qing-Hua Shen}~\email{shenqinghua@impcas.ac.cn}
\affiliation{State Key Laboratory of Heavy Ion Science and Technology, Institute of Modern Physics, Chinese Academy of Sciences, Lanzhou 730000, China} 
 \affiliation{School of Physical Science and Technology, Lanzhou University, Lanzhou 730000, China}
 \affiliation{School of Nuclear Sciences and Technology, University of Chinese Academy of Sciences, Beijing 101408, China}

\author{Jun-Xu Lu}~\email{ljxwohool@buaa.edu.cn}
\affiliation{School of Physics, Beihang University, Beijing 102206, China}

\author{Li-Sheng Geng}~\email{lisheng.geng@buaa.edu.cn}
\affiliation{School of Physics, Beihang University, Beijing 102206, China} \affiliation{Sino-French Carbon Neutrality Research Center, Ecole Centrale de Pekin/School of General Engineering, Beihang University, Beijing 100191, China} \affiliation{Peng Huanwu Collaborative Center for Research and Education, International Institute for Interdisciplinary and Frontiers, Beihang University, Beijing 100191, China} \affiliation{Beijing Key Laboratory of Advanced Nuclear Materials and Physics, Beihang University, Beijing 102206, China}
\affiliation{Southern Center for Nuclear-Science Theory (SCNT), Institute of Modern Physics, Chinese Academy of Sciences, Huizhou 516000, China}

\author{Xiang Liu}~\email{xiangliu@lzu.edu.cn}
\affiliation{School of Physical Science and Technology, Lanzhou University, Lanzhou 730000, China}
\affiliation{Lanzhou Center for Theoretical Physics, Key Laboratory of Theoretical Physics of Gansu Province, Key Laboratory of Quantum Theory and Applications of MoE, Gansu Provincial Research Center for Basic Disciplines of Quantum Physics, Lanzhou University, Lanzhou 730000, China} 
\affiliation{MoE Frontiers Science Center for Rare Isotopes, Lanzhou University, Lanzhou 730000, China}
 \affiliation{Research Center for Hadron and CSR Physics, Lanzhou University $\&$ Institute of Modern Physics of CAS, Lanzhou 730000, China}

\author{Ju-Jun Xie}~\email{xiejujun@impcas.ac.cn}
\affiliation{State Key Laboratory of Heavy Ion Science and Technology, Institute of Modern Physics, Chinese Academy of Sciences, Lanzhou 730000, China} 
\affiliation{School of Nuclear Sciences and Technology, University of Chinese Academy of Sciences, Beijing 101408, China}
\affiliation{Southern Center for Nuclear-Science Theory (SCNT), Institute of Modern Physics, Chinese Academy of Sciences, Huizhou 516000, China}
 
 \begin{abstract}
 
We present a theoretical investigation of the radiative decay process $\Omega(2012) \to \gamma \Omega$, where the $\Omega(2012)$ resonance with spin-parity $J^P=\frac{3}{2}^-$, is treated as a dynamically generated state from $\bar{K}\Xi(1530)$ and $\eta \Omega$ in $s$-wave and $\bar{K}\Xi$ in $d$-wave. The radiative decay width of the $\Omega(2012)$ is calculated using a triangular loop mechanism, where the $\Omega(2012)$ couples to the $\bar{K} \Xi(1530)$ channel. Subsequently, the final state interactions between $\Xi(1530)$ and $\bar{K}$ transition to a photon and $\Omega$ through the exchange of a $\Xi$ baryon. Our calculations yield a radiative decay width of $13.2 ^{+4.5}_{-5.6}$ keV, with uncertainties arising from the model parameters. This result provides valuable insights into the nature of the $\Omega(2012)$ resonance and its decay dynamics. It is expected that the calculations presented here could be verified by future experiments, which would open a new door for studying the still elusive nature of the $\Omega(2012)$. 

 \end{abstract}
\maketitle
\section{Introduction}

$\Omega$-like states, which contain at least three strange quarks, are particularly elusive. In 2018, the Belle Collaboration observed a new $\Omega$(-like) state, referred to as the $\Omega(2012)$, in the $\Xi^0 K^-$ and $\Xi^- K^0_S$ invariant mass distributions with a combined significance of 8.3$\sigma$~\cite{Belle:2018mqs}. Its $\Xi \pi \bar{K}$ and $\Xi \bar{K}$ decays were also studied by experiments later~\cite{Belle:2019zco,Belle:2021gtf,Belle:2022mrg}. The Belle collaboration concluded that the $\Omega(2012)$ resonance is most likely to have spin-parity $J^P = 3/2^-$. Recently, the BES\uppercase\expandafter{\romannumeral 3} Collaboration observed the $\Omega(2012)$ in $e^+e^-$ collisions with a significance of $3.5\sigma$, and they also observed a new $\Omega$(-like) state $\Omega(2109)$ with $4.1\sigma$~\cite{BESIII:2024eqk}. Simultaneously, the ALICE Collaboration  observed the $\Omega(2012)$ state in high-multiplicity-triggered $pp$ collisions at $\sqrt{s} = 13$ TeV, with a significance of $15\sigma$~\cite{ALICE:2025atb}. At present, the $\Omega(2012)$ is classified as a three-star state with unknown spin and negative parity in the Review of Particle Physics (RPP)~\cite{ParticleDataGroup:2024cfk}. Its average mass ($M_R$) and width ($\Gamma_R$) are $2012.5\pm 0.6$ MeV and $6.4^{+2.5}_{-2.0} \pm 1.6$ MeV, respectively~\cite{ParticleDataGroup:2024cfk}.

\begin{figure*}[htbp]
    \centering
    \includegraphics[scale=0.8]{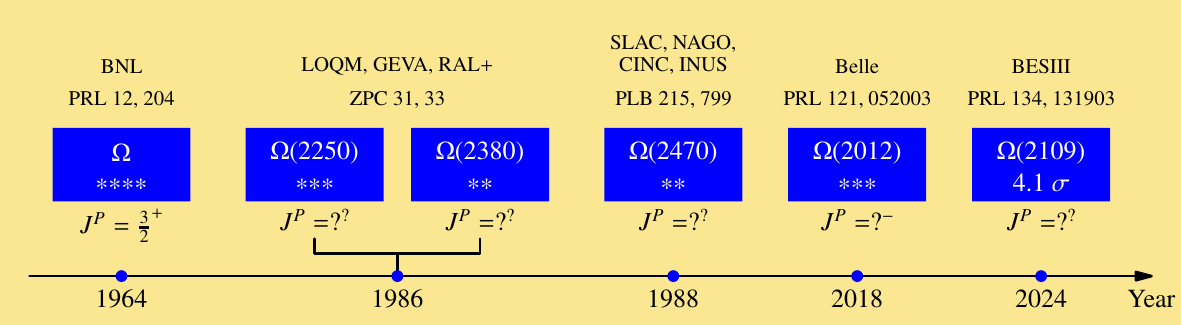}
    %\captionsetup{justification=raggedright, singlelinecheck=false}
    \caption{The current status of ground $\Omega$ and excited $\Omega^*$(-like) states. Data are taken from the RPP~\cite{ParticleDataGroup:2024cfk}.}
    \label{fig:status}
\end{figure*}

Before the observation of the $\Omega(2012)$ resonance, most information on $\Omega$(-like) states was extracted from bubble chamber experiments in the 1980s, and that information is very scarce~\cite {ParticleDataGroup:2024cfk}. There was only one three-star $\Omega$(-like) state, $\Omega(2250)$, with spin-parity unknown. Meanwhile, the other two higher two-star states, $\Omega(2380)$ and $\Omega(2470)$, are very poorly known, and they were omitted from the summary table in the RPP~\cite{ParticleDataGroup:2024cfk}. The current status of the ground $\Omega(1670)$ state and the four  $\Omega$-like resonances is shown in Fig.~\ref{fig:status}.

On the theoretical side, an $\Omega$ excited state with a mass around 2 GeV has been predicted by many theoretical works~\cite{Chao:1980em,Capstick:1986ter,Oset:1997it,Loring:2001ky,Pervin:2007wa,An:2014lga,Faustov:2015eba,Engel:2013ig,Kolomeitsev:2003kt,Sarkar:2004jh,Xu:2015bpl} before the $\Omega(2012)$ resonance was discovered. In particular, Ref.~\cite{Chao:1980em} reported a $J^P =3/2^-$ excited $\Omega$ resonance with a mass of about 2020 MeV within a quark model incorporating ingredients motivated by quantum chromodynamics. This state was found to decay into the $\bar{K}\Xi$ channel. Moreover, the radiative decay width of that $J^P =3/2^-$ excited $\Omega$ resonance with a mass of about 2020 MeV was calculated in Ref.~\cite{Kaxiras:1985zv} within the nonrelativistic potential model, yielding a value of $18$ keV. It is worth noting that the excited $\Omega$ sates with $J^P =1/2^-$ and $J^P =3/2^-$ in Refs.~\cite{Chao:1980em,Kaxiras:1985zv} are degenerate, sharing the same mass and partial decay width to the $\gamma \Omega$ channel.

After the observation of the $\Omega(2012)$ state, there have been a large variety of theoretical investigations about its nature (see more details in a short review of Ref.~\cite{Xie:2024wbd}). In the three-quark ($qqq$) baryon picture, various models find an $\Omega$ excited state near 2 GeV, such as the constituent quark model~\cite{Liu:2019wdr,Arifi:2022ntc,Menapara:2021vug,Wang:2022zja}, the harmonic oscillator expansion~\cite{Luo:2025cqs}, the QCD sum rule method~\cite{Su:2024lzy,Su:2025nfa}, and the lattice QCD method~\cite{Hudspith:2024kzk,Hockley:2024aym}. Meanwhile, the $\Omega(2012)$ is also interpreted as a molecular state considering that its mass is very close to the mass threshold of $\bar{K}$ and $\Xi(1530)$~\cite{Valderrama:2018bmv,Lin:2018nqd,Huang:2018wth,Pavao:2018xub,Polyakov:2018mow,Lin:2019tex,Ikeno:2020vqv,Hu:2022pae}. In the molecular picture, the strong decay modes of the $\Omega(2012)$ are the focus of studies, especially the $\Omega(2012) \to \bar{K}\Xi$ and $\bar{K}\Xi\pi$ processes. In Ref.~\cite{Lin:2018nqd}, the $\Omega(2012) \to \bar{K}\Xi$ and $\Omega(2012) \to \bar{K} \pi \Xi$ decays were investigated within the triangle loop diagram mechanism using the effective Lagrangian approach. It found that the dominant decay mode of the $\Omega(2012)$ state is the $\bar{K}\pi\Xi$ three-body channel. In Refs.~\cite{Lu:2020ste,Ikeno:2022jpe,Ikeno:2023wyh,Lu:2022puv,Zeng:2020och,Song:2024ejc}, the chiral unitary approach was employed to study the $\Omega(2012) \to \bar{K}\Xi\pi$ and $\Omega(2012) \to \bar{K}\Xi$ decays via the re-scattering of $\bar{K}\Xi(1530)$ and $\eta\Omega$ within the molecular framework. The current experimental results, such as the mass and width, can be naturally accommodated by the molecular nature of the $\Omega(2012)$.

\begin{figure}[htbp]
    \centering
    \includegraphics[scale=0.9]{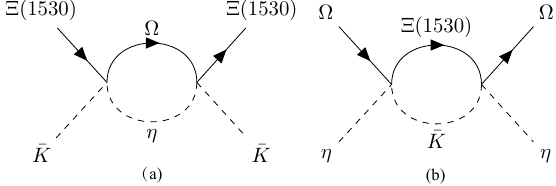}
    \caption{The interaction of (a) $\bar{K} \Xi(1530) \to \bar{K} \Xi(1530)$ and (b) $\eta\Omega \to \eta\Omega$ at the one-loop order.}
    \label{fig:production}
\end{figure}

It is worthy to mention that in the coupled-channel dynamics involving $\bar{K}\Xi(1530)$ and $\eta \Omega$, only the off-diagonal element does not vanish at leading order, i.e., $V_{\bar{K}\Xi(1530) \to \bar{K}\Xi(1530)}=V_{\eta \Omega \to \eta \Omega}=0$ within the chiral perturbation theory. Thus, the coupled-channel effects are crucial to dynamically generate the $\Omega(2012)$ state. The generation mechanism of the $\Omega(2012)$ from the $\bar{K}\Xi(1530) \to \bar{K} \Xi(1530)$ and $\eta\Omega \to \eta\Omega$ with this scenario at the one-loop order is displayed in Fig.~\ref{fig:production}. The molecular structure of $\Omega(2012)$ is rather peculiar: while it predominantly behaves as a bound state of $\bar{K}\Xi(1530)$, binding actually requires the coupling to the $\eta \Omega$ channel. Neither the $\bar{K}\Xi(1530)$ system nor the $\eta \Omega$ system would form a bound state on its own.

Recently, the structure of the $\Omega(2012)$ state was investigated by a combined analysis of lattice QCD simulations and experimental data within the Hamiltonian effective field theory~\cite{Han:2025gkp}, where the $\bar{K}\Xi$ and $\bar{K}\Xi(1530)$ channels were taken into account. It was found that the spin-parity of the $\Omega(2012)$ resonance is $3/2^-$, and the $\bar{K}\Xi(1530)$ channel plays a crucial role in shaping the structure and generating the real part of the $\Omega(2012)$ pole. Meanwhile, the $\bar{K} \Xi$ channel is responsible for the imaginary part of the pole, and thus contributes to the width of the $\Omega(2012)$.

Along this line, within the molecular picture, we study the radiative decay, $\Omega(2012) \to \gamma\Omega$, using the triangle loop mechanism in the present manuscript, where the $\Omega(2012)$ couples to the $\bar{K} \Xi(1530)$ channel. Subsequently, the final-state interactions between $\Xi(1530)$ and $\bar{K}$ transition to a photon and $\Omega$ through the exchange of a $\Xi$ baryon.

This paper is organized as follows. In Sec.~\ref{formalism}, we will explain the decay mechanism and calculate the radiative decay width for $\Omega(2012) \to \gamma \Omega$ via the triangle loop diagrams considering the strong coupling of the $\Omega(2012)$ to the $ \bar{K}\Xi(1530)$ channel. Numerical results and discussions are presented in Sec.~\ref{numericalresults}. Finally, a summary is given in Sec.~\ref{Summary}.

\section{Formalism and ingredients}~\label{formalism}

\begin{figure*}[htbp]
    \centering
    \includegraphics[scale=1.2]{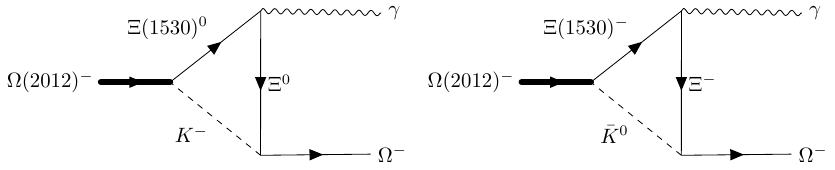}
    \caption{The radiative decay of $\Omega(2012) \to \gamma\Omega$ through triangle loop mechanism with the strong coupling of the $\Omega(2012)$ to the $\bar{K}\Xi(1530)$ channel.(a)$\Omega(2012)^-$ couples to $\Xi(1530)^0K^-$;(b)$\Omega(2012)^-$ couples to $\Xi(1530)^-\bar{K}^0$.}
    \label{fig:decay}
\end{figure*}

Considering the strong coupling of the $\Omega(2012)$ to the $\bar{K}\Xi(1530)$ channel, the dominant mechanism for the radiative decay $\Omega(2012) \to \gamma\Omega$ can proceed with the triangle loop mechanism shown in Fig.~\ref{fig:decay}, where the $\Omega(2012)$ directly couples to $\bar{K}\Xi(1530)$, then transit to $\gamma \Omega$ by exchanging a $\Xi$ hyperon. 

In the following, we provide a brief review of the unitary coupled-channel approach to investigate the $\Omega(2012)$ state. The effective interaction Lagrangians of $\Xi(1530)\Xi \gamma$ and $\Omega \bar{K} \Xi$ are also introduced.

\subsection{The molecular nature of $\Omega(2012)$ and its coupling to $\bar{K} \Xi(1530)$}

Within the chiral unitary approach, the coupled-channel interactions of $\bar{K}\Xi(1530)$ and $\eta\Omega$ with strangeness $S=-3$ and isospin $I=0$  have been investigated in Refs.~\cite{Kolomeitsev:2003kt,Sarkar:2004jh,Xu:2015bpl}. An $\Omega$ bound state with spin $\frac{3}{2}^-$ is found. In Ref.~\cite{Lu:2020ste}, the two- and three-body strong decays of $\Omega(2012) \to \bar{K}\Xi$ and $\Omega(2012) \to \bar{K}\pi\Xi$ were further investigated from the molecular perspective where the $\Omega(2012)$ resonance is a dynamically generated state from the coupled-channel interactions of $\bar{K}\Xi(1530)$ and $\eta \Omega$ in $s$-wave and $\bar{K}\Xi$ in $d$-wave. Following Refs.~\cite{Pavao:2018xub,Lu:2020ste}, the tree-level transition potentials among $\bar{K}\Xi(1530)$, $\eta \Omega$, and $\bar{K}\Xi$ read,
\begin{eqnarray}
&& V_{\bar{K}\Xi(1530) \to \bar{K}\Xi(1530) } = V_{\eta \Omega \to \eta \Omega } = V_{\bar{K} \Xi \to \bar{K} \Xi} = 0 , \\
&& V_{\bar{K} \Xi(1530) \to \eta \Omega } = V_{\eta \Omega \to \bar{K} \Xi(1530) } = -\frac{3}{4f^2_\pi}(E_K + E_\eta), \\
&& V_{\bar{K} \Xi(1530) \to \bar{K} \Xi } = V_{\bar{K} \Xi \to \bar{K} \Xi(1530)} = \alpha q^2_{\Xi}, \\
&& V_{\eta \Omega \to \bar{K} \Xi } = V_{\bar{K} \Xi \to \eta \Omega} =  \beta q^2_{\Xi},
\end{eqnarray}
where we take the pion decay constant $f_\pi = 93$ MeV.  The model parameters $\alpha$ and $\beta$ are involved in the $\bar{K}\Xi(1530)\to \bar{K}\Xi$ and $\eta \Omega \to \bar{K} \Xi$ transition potentials, respectively. The quantities 
\begin{eqnarray}
E_K &=& \frac{s + m^2_{\bar{K}} - M^2_{\Xi(1530)}}{2\sqrt{s}},\\
E_\eta &=& \frac{s + m^2_\eta - M^2_{\Omega}}{2\sqrt{s}},\\
q_\Xi &=& \frac{\sqrt{[s-(m_{\bar{K}} + M_{\Xi})^2][s-(m_{\bar{K}} - M_{\Xi})^2]}}{2\sqrt{s}},
\end{eqnarray}
are the center of frame (c.m.) energies of incoming or outgoing mesons, respectively, and the c.m. momentum of $K\Xi$ with $\sqrt{s}$ the invariant mass. With the transition potentials given above, the unitarized scattering amplitude $T$ can be obtained by solving the Bethe-Salpeter equation:
\begin{eqnarray}
T = V + VGT = [1-VG]^{-1} V,
\end{eqnarray}
where $G$ is the loop function for each channel, and it can be regularized with a cutoff prescription. Typically, the cutoff parameter ($q_{\rm max}$) in the loop functions for the $\bar{K}\Xi(1530)$, $\eta \Omega$, and $\bar{K}\Xi$ channels are different. In this work, we adopt a common value $q_{\rm max}$ for all channels, which will be determined from the Belle collaboration experimental data. We refer to Ref.~\cite{Lu:2020ste} for additional details regarding these model parameters.

Taking into account the updated experimental result of the ratio of the branching fraction of the three-body decay to that of the two-body decay ${\cal R}^{\bar{K}\pi\Xi}_{\bar{K}\Xi} = 0.99 \pm 0.26 \pm 0.06$ measured by the Belle Collaboration~\cite{Belle:2022mrg} and the experimental results for the mass $M_R = 2012.5 \pm 0.9 $ MeV and width $\Gamma_R = 6.4 \pm 3.0$ MeV of the $\Omega(2012)$ resonance, we have re-determined the model parameters, as well as the pole position $(M_R, \Gamma_R)$ of the $\Omega(2012)$ resonance and its couplings to the involving channels, which are listed in Table~\ref{tab:parameters}. Note that we take the same $q_{\rm max}$ for the $\bar{K}\Xi(1530)$, $\eta \Omega$, and $\bar{K}\Xi$ channels.

From Table~\ref{tab:parameters}, one can see that the obtained resonance parameters, mass and width, of the $\Omega(2012)$ states are in agreement with the experimental results within uncertainties.

\begin{table*}[htbp]
    \centering
    \renewcommand\arraystretch{1.5}
    \tabcolsep=6pt
    \caption{Model parameters and pole position $(M_R, \Gamma_R)$ (in MeV) of the $\Omega(2012)$ and its couplings to the involving channels. Parameter $q_{\rm max}$ is in MeV, and $\alpha$ and $\beta$ are in $ 10^{-9} \,{\rm MeV}^{-3}$.}
    \begin{tabular}{cccccccc}
        \toprule[1pt]
        \toprule[1pt]
     $q_{\rm max}$ & $\alpha$  & $\beta$  & $[M_R, \Gamma_R]$  & $|g_{\Omega(2012)\bar{K} \Xi(1530)}|$ & $|g_{\Omega(2012) \eta \Omega}|$  & $|g_{\Omega(2012)\bar{K}\Xi}|$ & ${\cal R}^{\bar{K}\pi\Xi}_{\bar{K}\Xi}$ \\
 %     (MeV) & ($\times 10^{-8} \,{\rm MeV}^{-3}$) & ($\times 10^{-8} \,{\rm MeV}^{-3}$) & (MeV) & & & &  \\
         \hline
         800   & $-175\pm10$  & $316\pm22$  & $[2014.1\pm1.0, 3.4\pm0.3]$ & $2.07\pm0.02$ & $1.87\pm0.07$  & $0.21\pm0.02$ & $0.76\pm0.15$ \\
         850   & $-174\pm11$  & $371\pm28$  & $[2014.1\pm1.0, 3.4\pm0.2]$ & $2.09\pm0.02$ & $1.56\pm0.05$  & $0.20\pm0.02$ & $0.81\pm0.14$ \\
         900   & $-157\pm11$  & $372\pm30$  & $[2014.1\pm1.0, 3.4\pm0.2]$& $2.10\pm0.02$ & $1.35\pm0.05$  & $0.20\pm0.02$ & $0.81\pm0.14$ \\
         1000  & $-120\pm9$  & $328\pm28$  & $[2014.2\pm1.0, 3.4\pm0.2]$ & $2.10\pm0.03$ & $1.07\pm0.04$  & $0.21\pm0.02$ & $0.81\pm0.14$ \\
         1100  & $-91\pm6$   & $273\pm24$  & $[2014.2\pm1.0, 3.4\pm0.2]$& $2.09\pm0.03$ & $0.88\pm0.04$ & $0.20\pm0.02$ & $0.81\pm0.14$ \\
         1200  & $-69\pm5$   & $219\pm22$  & $[2014.2\pm1.0, 3.5\pm0.3]$ & $2.08\pm0.03$ & $0.73\pm0.05$ & $0.21\pm0.03$ & $0.78\pm0.16$ \\
        \bottomrule[1pt]
        \bottomrule[1pt]
    \end{tabular}
    \label{tab:parameters}
\end{table*}

The coupling of the $\Omega(2012)$ to each relevant channel $g_i$ ($i =\Omega(2012) \bar{K}\Xi(1530)$, $\Omega(2012)\eta \Omega$, $\Omega(2012)\bar{K}\Xi$) can be determined via the residues of the scattering amplitude $T_{ij}$  at the pole position, which reads,
\begin{equation}
    \begin{aligned}
        T_{ij}=\frac{g_i g_j}{\sqrt{s}-z_R},
    \end{aligned}
\end{equation}
where $z_R = M_R - i \Gamma_R/2$ denotes the $\Omega(2012)$ pole position. The theoretical results for these coupling constants are also shown in Table~\ref{tab:parameters} for each scenario of model parameters and $q_{\rm max}$. Considering the dominant radiative decay mechanism shown in Fig.~\ref{fig:decay}, we need the coupling constant of $g_{\Omega(2012)\bar{K}\Xi(1530)}$, and in the numerical study we take
\begin{eqnarray}
g_{\Omega(2012)K^- \Xi(1530)^0} &=& g_{\Omega(2012) \bar{K}^0 \Xi(1530)^-} \nonumber \\  
&=& -\frac{1}{\sqrt{2}} g_{\Omega(2012)\bar{K}\Xi(1530)}.
\end{eqnarray}
In the present notation, we employ the isospin doublets ($\bar{K}^0$, $-K^-$) and ($\Xi(1530)^0$, $\Xi(1530)^-$). The resulting $I = 0$ state of the $\Xi(1530)\bar{K}$ system is written as~\cite{ParticleDataGroup:2024cfk}
\begin{eqnarray}
         \ket{\Xi(1530)\bar{K},I=0} &=& -\sqrt{\frac{1}{2}}\ket{\Xi(1530)^0K^-} \nonumber \\
        && -\sqrt{\frac{1}{2}}\ket{\Xi(1530)^-\bar{K}^0}.
\end{eqnarray}

It is important to note that, in Ref. [38], only the mass and width of the $\Omega(2012)$ were fitted, while the upper limit $R^{\bar{K}\pi\Xi}_{\bar{K}\Xi} < 0.119$ was imposed merely as a constraint. In the present work, however, the updated experimental value $R^{\bar{K}\pi\Xi}_{\bar{K}\Xi} = 0.99 \pm 0.26 \pm 0.06$ is included as an additional data point in the fit. This change in the fitting inputs explains the differences between the theoretical parameters obtained here and those reported in Ref.~\cite{Lu:2020ste}. Moreover, the larger absolute values of $\alpha$ and $\beta$ lead to a larger coupling constant $g_{\Omega(2012)\bar{K}\Xi(1530)}$ and a smaller $g_{\Omega(2012)\bar{K}\Xi}$, which in turn allows the new, larger experimental ratio ${\cal R}^{\bar{K}\pi\Xi}_{\bar{K}\Xi}$ to be reproduced within uncertainties. These results further support the molecular interpretation of the $\Omega(2012)$ state.

\subsection{The effective Lagrangians for the $\Xi(1530)\Xi\gamma$ and $\Omega \bar{K} \Xi$ interaction}

%To calculate the radiative decay width of $\Omega(2012) \to \gamma \Omega$ with the triangel loop diagrams shown in Fig.~\ref{fig:decay}, we also need the effective interactions for $\Xi(1530) \Xi\gamma$ and $\Omega \bar{K} \Xi$ vertexes.

For the $\Xi(1530)\Xi\gamma$ vertex, the effective Lagrangian reads~\cite{Solano:1998ne,Zou:2002yy,Oh:2004wp,Nakayama:2006ty,Oh:2007jd,Xie:2013mua,Wei:2021qnc,Zhang:2021esc,Wei:2022nqp,Wang:2023ijp,Wei:2024lne}
\begin{eqnarray}
          \mathcal{L}_{\Xi(1530)\Xi\gamma} &=& i\frac{eg_1}{2m_N}F^{\mu\nu}\bar{\Xi}\gamma_\mu \gamma_5\Xi(1530)_\nu \nonumber \\
    && \!\!\! \!\!\!\! \!\!\! \!\!\! -\frac{eg_2}{(2m_N)^2}F^{\mu\nu}(\partial_\mu\bar{\Xi})\gamma_5\Xi(1530)_\nu + H.c. \, ,
    \label{eq:DBgamma}
\end{eqnarray}
where $F^{\mu\nu}=\partial^\mu A^\nu-\partial^\nu A^\mu$ is the conventional photon field strength tensor and $A^\mu$ refers to the photon field. The above effective Lagrangian describes the interactions among the octet baryon, decuplet baryon, and photon. It automatically satisfies the gauge symmetry invariance~\cite{Xie:2013mua}. There are two electromagnetic transition coupling constants $g_1$ and $g_2$ in Eq.~(\ref{eq:DBgamma}), in which the $g_1$ term is dominant. In general, these two couplings can be determined from the radiative decay of $\Xi(1530) \to \gamma \Xi$. Yet, the experimental measurement of the radiative decay of $\Xi(1530) \to \gamma \Xi$ is currently insufficient~\cite{ParticleDataGroup:2024cfk}. Only theoretical calculations are available. Here, we take $g_1=3.02$ and $g_2=-2.40$ for the $\Xi(1530)^0 \Xi^0 \gamma$ vertex as shown in Fig.~\ref{fig:decay} (a), and $g_1=0.56$ and $g_2=-0.16$ for the $\Xi(1530)^- \Xi^- \gamma$ vertex as shown in Fig.~\ref{fig:decay} (b). The above values obtained via the chiral constituent quark model~\cite{Wagner:1998bu} are employed in Ref.~\cite{Nakayama:2006ty} for the study of photoproduction $\gamma N \to KK\Xi$. Note that the coupling strength of $\Xi(1530)^- \Xi^- \gamma $ is much weaker than $\Xi(1530)^0 \Xi^0 \gamma$ due to the SU$_F(3)$ flavor symmetry breaking effect. The decay width of $\Xi(1530)^- \to \Xi^-\gamma$ will be zero if SU$_F(3)$ flavor symmetry is exact~\cite{Wagner:1998bu}.

For the $\Omega \bar{K} \Xi$ vertex, we use the effective Lagrangian as used in previous works~\cite{Machleidt:1987hj,Sasaki:2006cx,Gao:2010ve,Ronchen:2012eg,Haidenbauer:2017sws,Yang:2018idi,Wang:2022vln}
\begin{equation}
    \begin{aligned}
        \mathcal{L}_{\Omega \bar{K} \Xi}=\frac{f_{\Omega\Xi\bar{K}}}{m_K} \bar{\Omega}_v \Xi \partial^v\bar{K} \, ,
    \end{aligned}
    \label{eq:DBP}
\end{equation}
where we take the coupling constant $f_{\Omega\Xi \bar{K}} = f_{\Omega^- \Xi^0 K^-}=-f_{\Omega^- \Xi^- \bar{K}^0} =  8.13$~\cite{Yang:2018idi}. This coupling constant can be obtained using the $SU(3)$ relation as in Refs.~\cite{Nakayama:2006ty,Haidenbauer:2017sws}.

\subsection{The radiative decay $\Omega(2012) \to \gamma \Omega$}
With the above formalism, one can obtain the radiative decay amplitude of $\Omega(2012) \to \gamma\Omega$ corresponding to the triangle loop diagrams shown in Fig.~\ref{fig:decay}: 
\begin{equation}
    \begin{aligned}
        \mathcal{M}_{\Omega(2012) \to \gamma \Omega} &=\int\frac{d^4q}{(2\pi)^4} \frac{\widetilde{A}F^2(q_{ex})}{q^2-m_{\Xi(1530)}^2+i\varepsilon} \times \\ 
        & \frac{1}{(P-q)^2-m_{K}^2+i\varepsilon}\frac{1}{q_{ex}^2-m_{\Xi}^2+i\varepsilon} ,
    \end{aligned}
    \label{eq:M}
\end{equation}
where $P$, $q$, and $q_{ex}$ are the four-momenta of the $\Omega(2012)$, $\Xi(1530)$, and the exchanged hyperon $\Xi$, respectively. $m_{\Xi(1530)}$ is the mass of $\Xi(1530)$, and
\begin{eqnarray}
\widetilde{A} &=& A_1 (\slashed{q}-\slashed{k_1}+m_\Xi) A_2^{\beta} \widetilde{B}_{\beta\nu}A_3^\nu , \label{eq:t} \\
A_1 &=& -i \frac{f_{\Omega\Xi K}}{m_K}\bar{u}_\mu (k_2,s_{\Omega})(P-q)^\mu , \\
A_2^\beta &=& \left\{-\frac{eg_1}{2m_N}\left[\slashed{k}_1g^{\alpha \beta}-k_1^\beta \gamma^\alpha\right] + \nonumber \right.\\
       &&\left.\frac{eg_2}{(2m_N)^2}\left[q \cdot k_1 g^{\alpha \beta}-k_1^{\beta}q^{\alpha}\right] \right\} \varepsilon_{\alpha}^*(k_1,\lambda) \gamma_5 , \\
\widetilde{B}_{\beta\nu} & = & (\slashed{q}+m_{\Xi(1530)})\left[-g_{\beta\nu}+\frac{1}{3}\gamma_{\beta}\gamma_{\nu}+\nonumber \right. \\ &&\left. \frac{2q_{\beta} q_{\nu}}{3m_{\Xi(1530)}^2}+\frac{\gamma_{\beta}q_{\nu}-\gamma_{\nu}q_{\beta}}{3m_{\Xi(1530)}} \right], \\
A_3^{\nu} &=& \frac{g_{\Omega(2012)\bar{K}\Xi(1530)}}{\sqrt{2}}u^\nu(P,s_{\Omega(2012)}).
\end{eqnarray}
Here, $k_1$ ($\lambda$) and $k_2$ ($s_{\Omega}$) are the four-momenta (spin polarization) of the final photon and $\Omega$ baryon. $u^\mu(P,s_{\Omega(2012)})$ is a Rarita-Schwinger spinor for the $\Omega(2012)$ resonance with momentum $P$ and spin polarization $s_{\Omega(2012)}$. The expression for the photon polarization vector $\varepsilon^\mu(k_1,\lambda)$ are given by
\begin{equation}
    \begin{aligned}
        \varepsilon^\mu(k_1, \pm 1)=\frac{\mp 1}{\sqrt{2}}
        \begin{pmatrix}
          0\\
          1\\
          \pm i\\
          0
        \end{pmatrix}
    \end{aligned}
    \label{eq:epsph}.
\end{equation}
And for the Rarita-Schwinger spinor, it can be written as
\begin{equation}
    \begin{aligned}
        u^{\mu}(p,s)=\sum_{r}\langle 1,r,\frac{1}{2},s-r|\frac{3}{2},s \rangle \varepsilon(p,r)u(p,s-r),
    \end{aligned}
\end{equation}
where $\langle 1,r,\frac{1}{2},s-r|\frac{3}{2},s \rangle$ is Clebsch-Gordan coefficient. More explicitly, 
\begin{eqnarray}
        u^{\mu}(p,+\frac{3}{2}) &=& \varepsilon(p,+1)u(p,+\frac{1}{2}),  \nonumber \\
        u^{\mu}(p,+\frac{1}{2}) &=& \sqrt{\frac{2}{3}}\varepsilon(p,0)u(p,+\frac{1}{2})+\sqrt{\frac{1}{3}}\varepsilon(p,+1)u(p,-\frac{1}{2}), \nonumber \\
        u^{\mu}(p,-\frac{1}{2}) &=& \sqrt{\frac{1}{3}}\varepsilon(p,-1)u(p,+\frac{1}{2})+\sqrt{\frac{2}{3}}\varepsilon(p,0)u(p,-\frac{1}{2}), \nonumber \\
        u^{\mu}(p,-\frac{3}{2}) &=& \varepsilon(p,-1)u(p,-\frac{1}{2}), \nonumber
\end{eqnarray}
where the spin-1 polarization vector $ \varepsilon(\vec{p},\lambda)$ and spin-$\frac{1}{2}$ Dirac spinor $u(p,\lambda)$ are:
\begin{eqnarray}
\varepsilon(\vec{p},\lambda) &=& \left( \frac{\vec{p} \cdot \vec{\epsilon}_{\lambda}}{m}, \vec{\epsilon}_{\lambda}+\frac{\vec{p} \cdot \vec{\epsilon}_{\lambda}}{m(E+m)}\vec{p} \right),\\
u(p,+\frac{1}{2}) &=&
        \begin{pmatrix}
            \sqrt{E+m}\\
            0\\
            \frac{p_z}{\sqrt{E+m}}\\
            \frac{p_{+}}{\sqrt{E+m}}\\
        \end{pmatrix}, \\
u(p,-\frac{1}{2}) &=&
        \begin{pmatrix}
            0\\
            \sqrt{E+m}\\
             \frac{p_{-}}{\sqrt{E+m}}\\
            -\frac{p_z}{\sqrt{E+m}}\\
        \end{pmatrix},
\end{eqnarray}
with $E = \sqrt{|\vec{p}|^2+m^2}$ and $p_{\pm}=p_x \pm ip_y$. And the $\vec{\epsilon}_{\lambda}$ are:
\begin{equation}
    \begin{aligned}
        \epsilon_0=
        \begin{pmatrix}
        0\\
        0\\
        1
        \end{pmatrix},  \hspace{0.8cm}
         \epsilon_{\pm 1}=\frac{\mp 1}{\sqrt{2}}
         \begin{pmatrix}
             1\\
             \pm i\\
             0
         \end{pmatrix}
    \end{aligned}.
    \label{eq:eps}
\end{equation}

Next, we introduce a form factor $F(q_{ex})$ to suppress the off-shell contributions from the exchanged $\Xi$ baryon, which reads~\cite{Lin:2018nqd,Lin:2017mtz},
\begin{equation}
  F(q_{ex})=\frac{\Lambda^4}{(q_{ex}^2 - m^2_{\Xi})^2 + \Lambda^4}, \label{eq:ff}
\end{equation}
where $\Lambda$ is a cutoff parameter, and $q_{ex}=q-k_1$. We note that incorporating this form factor can eliminate the divergence of the loop integration in Eq.~\eqref{eq:M}.

For the numerical evaluation of Eq.~(\ref{eq:M}), we use the rest frame of the initial particle $\Omega(2012)$, thus
\begin{equation}
    \begin{gathered}
        P=(P^0,0,0,0) = (M_{\Omega(2012)},0,0,0), \\
        q=(q_0, \left\lvert \vec{q}\right\rvert\sin\theta\cos\phi, \left\lvert \vec{q}\right\rvert\sin\theta\sin\phi,\left\lvert \vec{q}\right\rvert\cos\theta),\\
        k_1=(E_{\gamma},0,0,- E_\gamma), k_2=(E_{\Omega},0,0,E_\gamma),
    \end{gathered}
\end{equation}
with
\begin{equation}
        E_{\gamma/\Omega} =\frac{M_{\Omega(2012)}^2 \mp m_{\Omega}^2}{2M_{\Omega(2012)}},
\end{equation}
where $M_{\Omega(2012)}$ is the mass of $\Omega(2012)$, $E_{\gamma}(E_\Omega)$ is the energy of the final photon ($\Omega$). We first integrate out the $q_0$ component utilizing the residue theorem~\cite{Shen:2024jfr} and introduce a cutoff $\Lambda$ for the integration over $|\vec{q}|$~\footnote{ The triangle loop integral depends on the cutoff parameter $\Lambda$ from the form factor of the exchanged $\Xi$ in Eq.~\eqref{eq:ff}. We have therefore set the value of $\Lambda$ in this integral to be the same as that in the form factor. This treatment not only simplifies the calculation but also reduces the number of free parameters in the theory.},

\begin{equation}
   \begin{aligned}
      \mathcal{M}_{\Omega(2012) \to \gamma \Omega}=-\frac{i}{(2\pi)^3}\int^{\Lambda}_0 d^3\vec{q}  \frac{N}{D},
   \end{aligned}
\end{equation}

where 
\begin{eqnarray}
      D &=& 8\omega_{\Xi(1530)} \omega_{K}\omega_{\Xi}(P^0+\omega_{\Xi(1530)}+\omega_{K}) \nonumber \\
      & &\times (P^0-\omega_{\Xi(1530)}+\omega_{K}+i\epsilon) \nonumber \\
      & &\times (P^0-\omega_{\Xi(1530)}-\omega_{K}+i\epsilon)(E_{\gamma}+\omega_{\Xi(1530)}+\omega_{\Xi})\nonumber \\
      & &\times (E_{\gamma}-\omega_{\Xi(1530)}+\omega_{\Xi}+i\epsilon)(E_{\gamma}-\omega_{\Xi(1530)}-\omega_{\Xi}+i\epsilon) \nonumber \\
      & &\times (E_{\Omega}+\omega_{K}+\omega_{\Xi})(E_{\Omega}+\omega_{K}-\omega_{\Xi}+i\epsilon) \nonumber \\
      & &\times (E_{\Omega}-\omega_{K}-\omega_{\Xi}+i\epsilon), \\
      N &=& N_1+N_2+N_3,
\end{eqnarray}
with
\begin{eqnarray}
N_1 &=& 4\omega_{K}\omega_{\Xi}(P^0+\omega_{\Xi(1530)}+\omega_{K})(E_{\gamma}+\omega_{\Xi(1530)}+\omega_{\Xi}) \nonumber\\
      && \times (E_{\Omega}+\omega_{K}+\omega_{\Xi})(E_{\Omega}+\omega_{K}-\omega_{\Xi}) \nonumber \\
      && \times (E_{\Omega}-\omega_{K}-\omega_{\Xi})\widetilde{t}|_{\omega_{\Xi(1530)}}, \nonumber \\
N_2 &=&4\omega_{\Xi(1530)} \omega_{\Xi}(P^0-\omega_{\Xi(1530)}+\omega_{K})(E_{\gamma}+\omega_{\Xi(1530)}+\omega_{\Xi}) \nonumber \\
      && \times (E_{\gamma}-\omega_{\Xi(1530)}+\omega_{\Xi})(E_{\gamma}-\omega_{\Xi(1530)}-\omega_{\Xi})\nonumber \\
      && \times (E_{\Omega}-\omega_{K}-\omega_{\Xi}) \widetilde{t}|_{P_0+\omega_{K}}, \nonumber \\
  N_3 &=& 4\omega_{\Xi(1530)} \omega_{K}(P^0+\omega_{\Xi(1530)}+\omega_{K})(P^0-\omega_{\Xi(1530)}+\omega_{K})  \nonumber \\
      && \times (P^0-\omega_{\Xi(1530)}-\omega_{K})(E_{\gamma}-\omega_{\Xi(1530)}-\omega_{\Xi}) \nonumber \\
      && \times (E_{\Omega}+\omega_{K}+\omega_{\Xi})\widetilde{t}|_{E_{\gamma} + \omega_{\Xi}}, \nonumber
\end{eqnarray}
\noindent
where $\widetilde{t}|_x$ is the value of $\widetilde{t}$ obtained at $q^0=x$ with $\widetilde{t}=\widetilde{A}F^2(q_{ex})$, and
\begin{equation}
   \begin{aligned}
      \omega_{\Xi(1530)}&=\sqrt{|\vec{q}|^2+m_{\Xi(1530)}^2}, ~~~~~  \omega_{K} = \sqrt{|\vec{q}|^2+m_{K}^2},\\
      \omega_{\Xi}&=\sqrt{|\vec{q}-\vec{k_1}|^2+m_{\Xi}^2}.
   \end{aligned}
\end{equation}

With the expressions specified above, one can obtain the decay width as 
\begin{equation}
\Gamma_{\Omega(2012)\to\gamma\Omega}=\frac{E_\gamma}{32\pi M_{\Omega(2012)}^2}\overline{\sum_{s_{\Omega(2012)}}} \sum_{\lambda,s_{\Omega}}|\mathcal{M}_{\Omega(2012)\to \gamma \Omega}|^2  ,
    \label{eq:W}
\end{equation}
where we average over the spin of the initial state $\Omega(2012)$ and sum over the spins of final states $\gamma$ and $\Omega$.

\section{Numerical results}~\label{numericalresults}

We have one free parameter $\Lambda$ as introduced in the form factor of Eq.~\eqref{eq:ff}. Here, we take its value in the range of $850 \sim 1100$ MeV as one of the $q_{\rm max}$ we have determined in Table~\ref{tab:parameters}. When $q_{\rm max}$ is in that range, the obtained pole position of the $\Omega(2012)$, the coupling constant $|g_{\Omega(2012)\bar{K}\Xi(1530)}|$, and the ratio ${\cal R}^{\bar{K}\pi \Xi}_{\bar{K}\Xi}$ are stable.

\begin{figure}[htbp]
    \centering
    \includegraphics[scale=0.85]{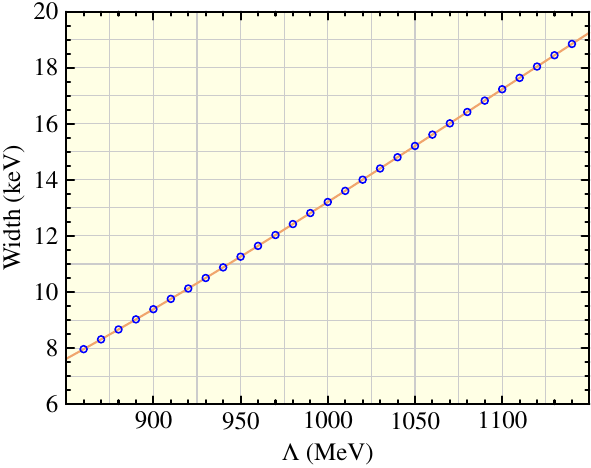}
    \caption{The obtained partial decay width of $\Omega(2012)$ $\to \gamma \Omega$ as a function of the parameter $\Lambda$.}
    \label{fig:width}
\end{figure}

With the coupling constant $|g_{\Omega(2012)\bar{K}\Xi(1530)}| \approx 2.1$ in Table~\ref{tab:parameters} and formalism in the previous section, we can obtain the radiative decay width of $\Omega(2012) \to \gamma\Omega$. The radiative decay width as a function of the parameter $\Lambda$ is shown in Fig.~\ref{fig:width}. In this work, the physical masses of the involved particles are taken from the RPP~\cite{ParticleDataGroup:2024cfk} and we have tabulated them in Table~\ref{tab:mass}.

\begin{table}[htbp]
    \centering
    \renewcommand\arraystretch{1.5}
    \tabcolsep=6pt
    \caption{Masses (in MeV) of the particles used in this work.}
    \begin{tabular}{ccccc}
        \toprule[1pt]
        \toprule[1pt]
        $\Xi(1530)^-$ & $K^-$ & $\Xi^-$ & $\Omega^-$ & $\Omega(2012)^-$\\
        \hline
         1535.0&493.677 &1321.71 &1672.45 &2012.4\\
         \hline
         $\Xi(1530)^0$ & $\bar{K}^0$ & $\Xi^0$ &$N$ &\\
         \hline
          1531.8&497.611 &1314.86 &938.3 & \\
        \bottomrule[1pt]
        \bottomrule[1pt]
    \end{tabular}
    \label{tab:mass}
\end{table}

Within the range of $\Lambda = 1000^{+100}_{-150}$ MeV, we obtain $\Gamma_{\Omega(2012) \to \gamma \Omega} = 13.2 ^{+4.5}_{-5.6}$ keV. Combining this partial decay width with the theoretical total width $\Gamma_R = 3.4\pm 0.3$ MeV of the $\Omega(2012)$ resonance, one can estimate the branching fraction of this
radiative decay process:
\begin{equation}
BR[\Omega(2012) \to \gamma \Omega] = 3.9^{+1.4}_{-1.7} \times 10^{-3}.
\end{equation}

Apart from the molecular picture, the radiative decay width $\Gamma_{\Omega(2012) \to \gamma \Omega}$ has also been calculated in the quark model~\cite{Liu:2019wdr}, which is about $9.52$ keV. This result was derived on the basis that the $\Omega(2012)$ resonance is assigned to a $3/2^-$ excited 1P-wave $\Omega$ excited state. The partial decay width obtained here is consistent with the one obtained in Ref.~\cite{Liu:2019wdr} within uncertainties. However, it should be noted that the branching fraction obtained here, $BR[\Omega(2012)\to\gamma\Omega]$, differs from the value ($1.67\times10^{-3}$) given in Ref.~\cite{Liu:2019wdr}, where a total width $\Gamma_R=5.69$ MeV was used. In this work, we employ $\Gamma_R = 3.4\pm 0.3$ MeV, as listed in Table~I. Uncertainties arising from the errors in its coupling constant to the $\bar{K}\Xi(1530)$ channel are not included, since their contributions are small compared to the error originating from the range of the cutoff parameter $\Lambda$. We hope that this decay channel will be measured in future experiments, providing a test for the model calculations presented here.

Until now, the $\Omega(2012)$ has been observed by three collaborations---Belle, BES\uppercase\expandafter{\romannumeral 3}, and ALICE. Its mass has been determined, while its width has large uncertainties. Its strong decay modes $\Omega(2012) \to \bar{K}\Xi$ and $\bar{K}\Xi\pi$ are measured by Belle~\cite{Belle:2019zco,Belle:2021gtf,Belle:2022mrg} and ALICE~\cite{ALICE:2025atb}. The investigation of its electromagnetic interactions is still lacking. Furthermore, its spin and parity are not fully confirmed. These experimental measurements can be carried out not only by the Belle collaboration but also by the Belle II, LHCb, and BESIII collaborations in the near future, as well as by the next-generation facilities~\cite{KLF:2020gai,Aoki:2021cqa}.

\section{Summary}~\label{Summary}

The study of $\Omega$(-like) states, which contain at least three strange quarks, has long been challenging due to their scarcity. A significant advancement occurred in 2018 when the Belle Collaboration discovered the $\Omega(2012)$ resonance in the $\Xi^0 K^-$ and $\Xi^- K^0_S$ invariant mass distributions with a combined significance of $8.3\sigma$~\cite{Belle:2018mqs}. Subsequent analyses of its $\Xi \pi \bar{K}$ and $\Xi \bar{K}$ decay channels suggested its spin-parity is most likely $J^P = 3/2^-$. Recent confirmations have solidified its status: the BESIII Collaboration observed it in $e^+e^-$ collisions ($3.5\sigma$) and reported a new state, the $\Omega(2109)$ ($4.1\sigma$)~\cite{BESIII:2024eqk}, while the ALICE Collaboration observed it in $pp$ collisions with a high significance of $15\sigma$~\cite{ALICE:2025atb}. The RPP currently lists the $\Omega(2012)$ as a three-star state with unknown spin and negative parity, reporting an average mass and width of $M_R = 2012.5 \pm 0.6$ MeV and $\Gamma_R = 6.4^{+2.5}_{-2.0} \pm 1.6$ MeV, respectively~\cite{ParticleDataGroup:2024cfk}.

Prior to these discoveries, information on $\Omega$ excitations was scarce and primarily came from 1980s bubble chamber experiments, with the $\Omega(2250)$ being the only well-established three-star state. The observation of the $\Omega(2012)$ thus marks a pivotal breakthrough, opening a new chapter in the study of triply-strange baryons. However, the internal structure of the $\Omega(2012)$ remains controversial. Consequently, studying its production and decays in various processes is essential to gain crucial insights into its nature.

In this work, we investigate the radiative decay $\Omega(2012)\to\gamma\Omega$ within a molecular picture, where the $\Omega(2012)$ resonance with quantum numbers $J^P=\frac{3}{2}^-$, is dynamically generated by the coupled channel interactions of $\bar{K}\Xi(1530)$ and $\eta \Omega$ in $s$-wave and $\bar{K}\Xi$ in $d$-wave. By considering the dominant contribution from the triangle loop diagrams in which the $\Omega(2012)$ directly couples to $\bar{K}\Xi(1530)$ and then transitions to $\gamma \Omega$ via the exchange of a $\Xi$ hyperon, we have determined the partial decay width of $\Omega(2012) \to \gamma \Omega$ to be $13.2 ^{+4.5}_{-5.6}$ keV. This result offers valuable information on the nature of the $\Omega(2012)$ resonance and its decay dynamics. We anticipate that this result obtained here can be tested in future experiments by the Belle \uppercase\expandafter{\romannumeral 2} and/or BES\uppercase\expandafter{\romannumeral 3} collaborations, and such experimental measurements can be used to study the nature of the $\Omega(2012)$ state in great detail.

\begin{acknowledgments}

We want to thank Profs. Cheng-Ping Shen and Xian-Hui Zhong for useful discussions. This work is partly supported by the National Key R\&D Program of China under Grant No. 2023YFA1606703, and by the National Natural Science Foundation of China under Grant Nos. 12435007, 12361141819, and 1252200936. X.L. is also supported by the National Natural Science Foundation of China under Grant Nos. 12335001 and 12247101, the ‘111 Center’ under Grant No. B20063, the Natural Science Foundation of Gansu Province (No. 22JR5RA389, No. 25JRRA799), the fundamental Research Funds for the Central Universities, the project for top-notch innovative talents of Gansu province, and Lanzhou City High-Level Talent Funding.
 
\end{acknowledgments} 

\section*{DATA AVAILABILITY}
The data support the findings of this article are not publicly available. The data are available from the authors upon reasonable request.

\normalem
\bibliographystyle{apsrev4-1.bst}
\bibliography{reference.bib}
\end{document}